\documentstyle[12pt,a4,epsf]{article}
\def\be{\begin{equation}}
\def\ee{\end{equation}}
\def\beq{\begin{eqnarray}}
\def\eeq{\end{eqnarray}}

\begin{document}
\begin{flushright}
TIFR/TH/01-31
\end{flushright}
\medskip
\begin{center}
{\large\underbar{\bf Percolation Systems away from the Critical Point }}
\\[1cm]
Deepak Dhar \\ 
%\vspace{1cm}
Tata Institute of Fundamental Research,\\ 
Homi Bhabha Road, Mumbai 400 005, INDIA \\ 
e-mail: ddhar@theory.tifr.res.in
%\vspace{1cm}
\end{center}
\begin{center}
\underbar{\bf ABSTRACT}
\end{center}
%\medskip
   This article reviews some effects of disorder in percolation systems
even away from the critical density $p_c$. For densities below $p_c$, the
statistics of large clusters defines the animals problem. Its relation to
the directed animals problem and the Lee-Yang edge singularity problem is
described. Rare compact clusters give rise to Griffiths singuraties in the
free energy of diluted ferromagnets, and lead to a very slow relaxation of 
magnetization. In biassed diffusion on percolation clusters, trapping in
dead-end branches leads to asymptotic drift velocity becoming zero for
strong bias, and very slow relaxation of velocity near the critical bias
field.  
    
\medskip

\section{ Introduction}
 
  Let me start by thanking Professor H. R. Krishnamurthy and other members
of the organizing committee for inviting me to this meeting to felicitate
Professor Narendra Kumar on his 60th birthday, and giving me an
opportunity to pay my tribute to him.

  Over years, I have always enjoyed discussing various questions with
Narendra. The large spectrum of his interests, and his spirit of enquiry,
and insights to disentangle the essential problem from confusing
camouflage have been a source of admiration for me. Thus I am really very
happy to come here, and express my respect for him, and join the other
speakers in wishing him many more years of happy questing.

  While I have shared with Kumar a common interest in understanding
disordered systems, my own work has been largely in classical statistical
mechanics ($\hbar =0 $), while Kumar's contributions to quantum problems
have been discussed by several speakers here. Even so, I can legitimately
claim to be one of Kumar's coworkers. Our only paper together was on the
behavior of $\pm J$ spin-glass on a Bethe lattice, and was presented at
the DAE Solid State Physics Symposium meeting at Madras ( now Chennai) in
December, 1979. It turned out that neither Narendra nor I could attend the
meeting, so the paper was presented by our host, Profesor Krishnamurthy.
The abstract of the paper appeared in the proceedings of the symposium.
Unfortunately, both of us got enmeshed in other problems, and did not ever
write up this work for publication in regular journals.

   Professor Krishnamurthy has suggested that it would be better if I
discuss generally the different effects of disorder in classical
statistical mechanics, and not narrowly focus on the topic of my latest
preprint. I was rather relieved by this, as my paper-writing has been even
slower than my usual (slow) rate in the recent past.  So, in this talk, I
will not describe any new results, but discuss some examples of
interesting effects caused by disorder in a classical percolation system.
The selection of topics was determined primarily by my familiarity rather
than any other reasons. I will only outline the main results. The
interested reader will have to go back to the cited literature for
details.

  Consider a random mixture of equal-sized conducting and insulating beads
in a box. It is easy to see that as the fraction of conducting balls is
varied from zero to one, the bulk mixture undergoes a transition from
insulating to conducting. This is the basic percolation transition, and
has been studied  a lot as a simple geometrical model of phase
transitions. As in the case of thermal critical phase transitions, various
physical quantities have singular behavior near the phase transition,
characterized by "critical exponents". Scaling theory, simulations and
theoretical techniques such as renormalization group etc. have been used
very successfully to understand the behavior of percolation systems at its
critical point. In particular, in two dimensions, all the critical
exponents of percolation are known from conformal field theory. Several
good reviews of this subject are now available \cite{percoreviews}.

  There is more to percolation theory than the critical exponents.  Of
course, an experimental disordered system may often be modelled by site-
or bond- percolation. In general, one is much more likely to find it not
near its percolation threshold.  Critical exponents of the percolation
theory are not of much use in describing these systems. It turns out that
percolation systems, as classical models of disordered media show many
interesting features, even away from the critical point. I will discuss
some examples of these in the following.

\section{ Off-critical Exponents in Percolation}

  Let $P_n(p)$ be be probability that the
cluster of connected sites containing the origin has exactly $n$ sites.
Then,  for all $ p < p_c$, for large $n$,
\begin{equation}
P_n(p) \sim A n^{-\theta} exp[ - B(p) n]
\label{E:Pn}
\end{equation}
where B(p) is  a p-dependent function that goes to zero as $p$ tends to
$p_c$. The exponent $\theta$ is independent of $p$, and depends only on
the dimension of space. This law is valid for $n$ much greater than the
typical cluster size $ n^{\star}(p)$. For $n \ll n^{\star}(p)$, one gets
a different exponent $ P_n(p) \sim n^{-\tau}$, with $\tau \neq \theta$. 
As $p$ tends to $p_c$, $n^{\star}(p)$ diverges. Here $\tau$ is a critical
exponent of percolation theory, but $\theta$ is an {\it off-critical}
exponent.

 In the limit $p \rightarrow 0$, all clusters of
$n$ sites have the equal weight ( $ p^n$). Let $A_n$ be the number of  
different clusters of $n$ sites possible that contain the origin. (These
are  called rooted animals: same cluster with different positions of
origin are counted  as distinct \cite{rooted}.) Then
one expects that for large $n$,
\begin{equation}
A_n \simeq K \lambda^n n^{-\theta}
\end{equation}
The exponent $\theta$  in this equation is same as that defined by
Eq.(\ref{E:Pn}), because for small $p$, $P_n(p) \sim A_n p^n$.  One can
also define the average linear size of an animal of $N$ sites. This
grows as $n^{\nu}$, where the exponent $\nu$ is related to the exponent
$\theta$ defined above by the relation
\begin{equation}
\theta = ( d-2) \nu
\end{equation}
where $d$ is the dimension of space. The above equation is valid for $1
\leq d \leq 8$. For $d \geq 8$, the exponents $\theta $ and $\nu$ stick to
their mean-field values $3/2$ and $1/4$ respectively.  Eq.(3)
has the form of a hyperscaling relation except that $(d-2)$ appears here
instead of $d$. This is understood as being due to a hidden supersymmetry
in the problem \cite{parisi}, which makes the problem of determining the
number of animals in $d$-dimensions related to the problem of Lee-Yang
edge singularity in $(d-2)$ dimension.

  The Lee-Yang description of the mathematical mechanism of phase
transitions is well-known \cite{leeyang}. For a hard-core lattice gas in a
finite volume, with possible additional attractive short-ranged
interactions, the grand-canonical partition function is a finite degree
polynomial with positive coefficients in the chemical activity $z$. The
zeroes of this polynomial often ( not always) lie on lines in the
complex-$z$ plane. As the temperature is varied, the coefficients of the
polynomial change, and the zeroes move. If at some temperature, the zeroes
come arbitrarily close to the real axis as the size of the system is
increased, the free energy per site becomes a non-analytic function of
$z$, signalling the onset of a phase-transition.

  The density of zeroes along such a line of zeroes near  its end point
shows a power-law dependence on the distance from the endpoint. We define
the Lee-Yang edge singularity exponent $\sigma$ by the relation that the 
density varies as $\epsilon^{\sigma}$ at distance $\epsilon$ from the
endpoint. It turns out that $\sigma$ is independent of temperature for
all temperatures $T$ above the criticial temperature $T_c$, and
depends only on the dimension of the system. It was shown by Parisi and
Sourlas \cite{parisi} that the animal exponent $\theta(d)$ in
$d$-dimensions is related to the Lee-Yang edge singularity exponent in
$d-2$ dimensions 
\begin{equation}
\theta(d) = \sigma(d-2) + 1
\end{equation}

 If we allow only neighbors in the ``forward direction'' , ( say along the
direction of increasing coordinates on a hypercubical lattice), we get
animals with  a directional constraint. It turns out \cite{day} that
critical exponents of directed animals in $d$ dimensions turn out to be
related to those of undirected animals in $(d+1)$ dimensions.
\begin{equation}
\theta_{dir}(d) = \theta(d+1)
\end{equation}
and the transverse size exponent $\nu_{\perp,dir}$ for directed
animals in $d$
dimensions is the same as the (only one) size exponent $\nu$
for undirected animals in $(d+1)$ dimensions
\begin{equation}
\nu_{\perp,dir}(d) = \nu(d+1).
\end{equation}

 The exponent $\sigma$ is easily shown to take the values $\sigma = -1$
for $d=0$ ( a point), and $\sigma = -1/2$ for $d=1$. For $d=2$, one can
use the exact solution of hard hexagon lattice gas by Baxter to show that
$\sigma =-1/6$ for $d=2$ \cite{dd83}. This then shows that the exact
values of the exponent $\theta$ for undirected animal in dimensions $
1,2,3,4$ are $-1, 0, \frac{1}{2},\frac{5}{6}$ respectively. The
corresponding values of the size exponent $\nu $ are $1,
\frac{1}{2},\frac{5}{12}$ in dimensions $1, 3, 4$ respectively. In two
dimensions, the exponent $\nu$ is not determined by scaling relations
given above.  The upper critical dimension for the animals problem is $8$,
and for all $d \geq 8$, we get $\theta = \frac{3}{2}$, and $\nu =
\frac{1}{4}$.  The exponents for directed animals are easily determined
from the scaling relations given above.

  We have seen that there is a fairly good understanding of  the
off-critical ``below $p_c$'' exponents of
percolation.  One can also define off-critical exponents in the
super-critical regime of percolation theory. It was shown by Kunz and
Suillard \cite{kunz} that for all $p > p_c$, the probability that the
origin belongs to a finite cluster of $n$ sites varies as $exp( - b(p)
n^{\frac{d-1}{d}})$, for sufficiently large $s$. This is easy to
understand: to get a finite cluster of $n$ sites, we need to disconnect a 
it from the infinite cluster. This needs perimeter bonds order (
$n^{\frac{d-1}{d}}$)
bonds, which gives the result. More accurately, there is  apower-law
prefactor multiplying the exponential term, and  the probability for a
finite cluster of $n$ sites varies as 
\begin{equation}
P_n(p) \sim K n^{-\theta'} exp( - b(p) n^{\frac{d-1}{d}})
\end{equation}
where $b(p)$ is a function of $p$, and $K$ is a constant.
The exponent $\theta'$ can be calculated exactly \cite{lubensky} using the
fact that 
such clusters are roughly compact, with linear size varying as
$n^{\frac{1}{d}}$, and the fluctuations in the $(d-1)$-dimensional
roughly spherical surface can be described in terms of normal modes of
vibration of the surface. Lubensky and McKane using field theory
techniques showed that the exponent $\theta'$ takes the values
$\frac{5}{4},-\frac{1}{9},\frac{1}{8},-\frac{449}{450},-\frac{11}{12}$
for $d=2,3,4,5,6$ respectively. The
non-monotonic behavior of $\theta'$ as a function of $d$ comes from the
fact that the integrals over normal modes in odd and even dimensions
coming in the theory have different behaviors.   

Note that this is one of the few cases where non-trivial values of
exponents can be exactly calculated in many dimensions greater than 2.

\section{ Relaxation in Disordered Ferromagnets}

   Consider now the case when the atom at each occupied site in a
percolation network carries a magnetic moment, and there is a nearest
neighbor ferromagetic interaction $J$ between the magnetic atoms. The
``unoccupied'' sites may be vacancies, or occupied by nonmagnetic atoms.
Let
$p$ denote the concentration of magnetic atoms. Then if $p < p_c$, the
percolation threshold, then the system breaks up into
mutually-disconnected clusters of magnetically coupled spins. In such a
system no long-range spontaneous magnetization is possible at any non-zero
temperature. For $ p > p_c$, there is an infinite connected cluster, and
at sufficiently low temperatures, spontaneous magnetization exists. The
transition temperature $T_{curie}(p)$ depends on $p$, and goes to zero, as
$p$ is decreased from $1$ to $p_c$. The ferromagnetic phase is denoted 
by F in Fig. 1. 

   For $p < p_c$, there is no spontaneous magnetization. Let $f(T,h)$ be
the disorder-averaged free energy per site of this system at a temperature
$T$ in a magnetic field $h$.  However, it was shown by Griffiths
\cite{griffiths} that $f(T,h)$ is a non-analytic function of the magnetic
field at $h =0$ for all $ T < T_{curie}(p=1)$. While the partial
derivatives $\frac {\partial ^n }{\partial h^n} f(T,h)$ exist for all
positive integers $n$, and are finite, the Taylor series for $f(T,h)$ in
powers of $h$ does not converge for any $ T $ below the $T_{curie}(p=1)$,
the Curie-temperature of the ``pure'' system. Thus, in the entire region
marked G ( for the Griffiths phase) in fig. 1, there is no spontaneous
magnetization, but the free
energy per site $f(T,h)$ is a non-analytic function of $h$.

\begin{figure}
\centerline{
\epsfxsize=12.0cm
\epsfysize=9.0cm
\epsfbox{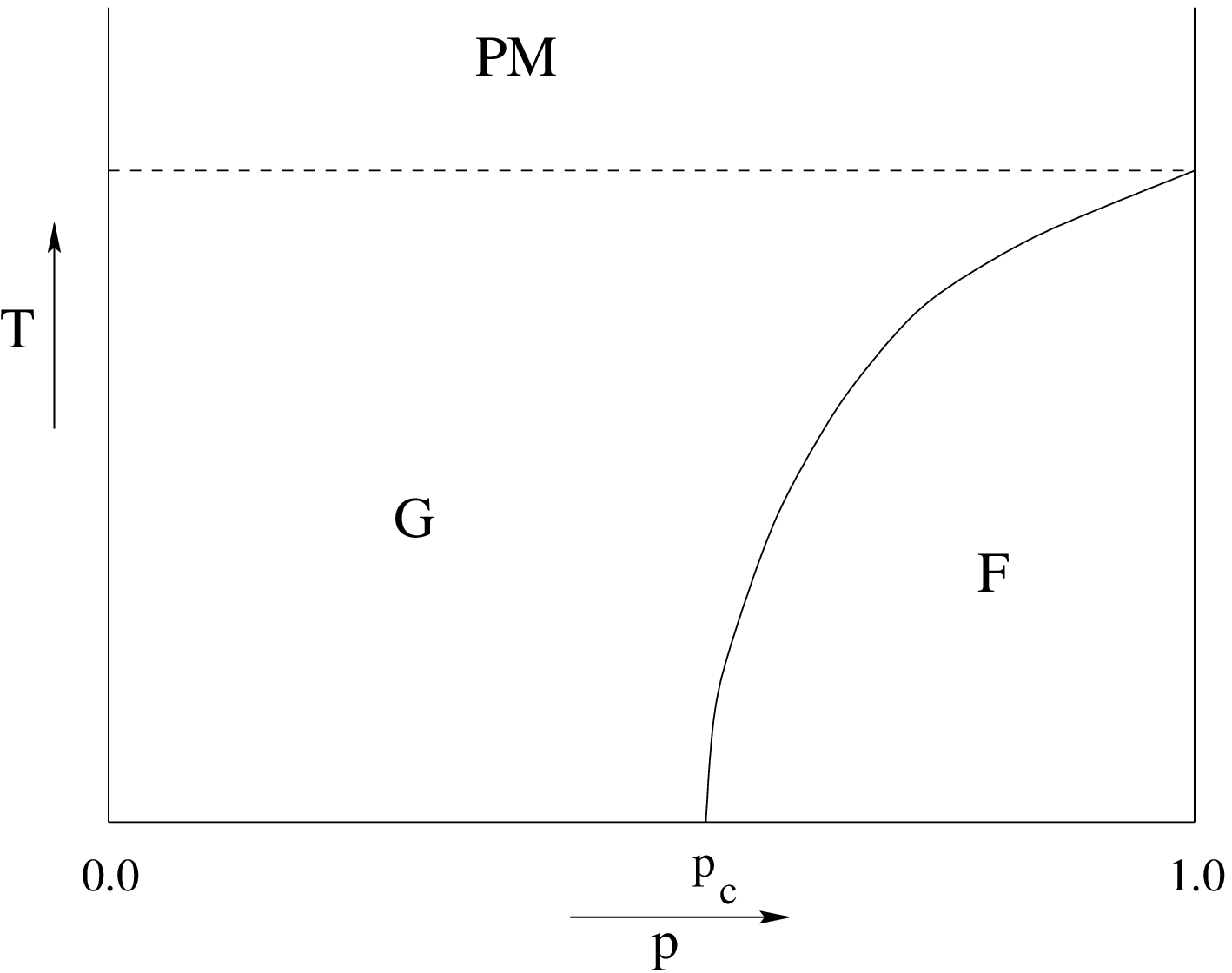}
}

\caption{}
\label{Fig.1}
\end{figure}

   While the 
nonanalyticity in $f(T,h)$ as  a function of $h$ is of the nature of  an
essential singularity, and is difficult to verify in experimental systems,
the rare large clusters responsible for it have  a much more pronounced
effect on the rate of relaxation to equilibrium in disordered systems.
Consider the decay of magnetization in such a system. We assume that the
system is coupled to heat bath at temperatute $T < T_{curie}(p=1)$, and
relaxes by single-spin -flip Glauber dynamics. At time $t=0$, the system
is prepared in a state with all spins up. We monitor the bulk
magnetization $M(t)$ at large times $t$.  Since the spins in different
clusters
do not interact with each other, we can write $M(t)$ as a weighted sum
over different cluster configurations ${\cal C}$.
\begin{equation}
M(t) = \sum_{C} {\rm Prob}({\cal C}) <S(t)>_{{\cal C}}
\label{E:grif}
\end{equation}
For a given finite cluster ${\cal C}$ of $n$-sites, one can determine the
average
magnetization at a site in the cluster by explicit diagonalization of the
of matrix of size $2^n \times 2^n$. This is non-trivial, except for very
small $n$. Fortunately, the behavior of $M(t)$ for large $t$ can be
determined by simple qualitative arguments.

In the summation Eq.(\ref{E:grif}), the leading behavior of each term is 
exponential in time, with the decay rate $\gamma_{{\cal C}}$ depends on
the cluster ${\cal C}$. Thus, we may write
\begin{equation}
<S(t)>_{\cal C} ~~\simeq ~~exp( - \gamma_{\cal C} ~t)
\end{equation}
and hence
\begin{equation}
M(t) \simeq  \sum_{C} {\rm Prob}({\cal C}) exp( - \gamma_{\cal C} ~t)  
\end{equation}

At large times, clusters with smallest decay rates
contribute most. The slowest relaxing clusters are those where all sites
within a disc of radius  $R$ are occupied. The  density of such clusters
varies as $exp( - A R^d)$, where the constant $A$ depends on $p$.  If the
magnetization of  such a cluster has to flip, it would need to create  a 
domain wall of energy $\simeq \sigma R^{d-1}$. The rate of such  
activated process will decrease for large $R$ as $exp ( - \sigma
R^{d-1}/T)$. It is then straight forward to put these estimates in the
Eq.(\ref{E:grif}), and deduce that
\begin{equation}
M(t) \sim exp( -K ( log t)^{\frac{d}{d-1}}),{\rm~~~~ for~~ large~~ t}.
\end{equation}
For $p>p_c$, in the ferromagnetic phase (F in fig. 1) the probability of
a large finite cluster of radius $R$, varies as $exp( - a R^{d-1})$, and
not
as $exp( - R^d)$. Then the steepest descent calculation shows that 
in the ferromagnetic phase $F$, the magnetization at long times decays as
a power law $M(t) \sim t^{-c}$, where the exponent $c$ depends on both $p$
and $T$.

The argument outlined above was first presented in \cite{ddstocha}. The
argument has been refined \cite{randeria,bray}. Unfortunately, neither
actual experimental data on disordered ferromagnets, nor results of
numerical simulations \cite{takano} show a clear evidence of such a $exp[
- (log t)^x]$ behavior. Presumably the time scale beyond which the
contribution of rare clusters will dominate is larger than experimentally
accessible time-scales. The experimental data seems to fit better a
stretched exponential $exp( - t^x)$. It seems that a more careful
argument, that gives not only the correct asymptotic behavior at longest
times, but also at intermediate times is needed.

\section{ Biassed Diffusion on Percolation Networks}

   Consider the motion of a single diffusing particle on a percolation
network, say in two dimensions, with density of occupied sites being $p$.
We assume that the
particle can move only on the occupied sites of the lattice. Then if $p$
is less than the critical probability $p_c$, the particle is localized.
For $p > p_c$, if the particle starts on the infinite cluster, its mean
square deviation from the initial position grows linearly with time,
$<R^2> \sim D(p) t$, where the diffusion constant $D(p)$ depends on $p$
and tends to zero as $p$ tends to $p_c$. This problem of (unbiassed)
diffusion on percolation clusters has been studied a lot \cite{unbiassed}.

   If there is a larger probability of displacement in some direction, due
to an imposed field, we have biassed diffusion. We shall model it by
assuming that at any time the diffusing particle attempts to take a step
in the up, right, down, left directions with probabilities $(1-B)/4, 1/4,
(1+B)/4$ and $1/4$ respectively. The step is actually taken if the
intended destination site is occupied. If the biassing field $B$ is small,
we have a non-zero value of average displacement per step, and this gives
rise to mean displacement in time $t$ in the direction of the field
growing linearly with $t$, and the mean velocity in time time $t$ tends to
a constant 

\begin{equation} 
\vec{v}_{\infty} = {\rm Lim}_{t \rightarrow \infty} <\vec{R}_t > /t 
\end{equation}

This asymptotic velocity $\vec{v}_{\infty}$ is
proportional to $B$ for small $B$.

If $p$ is near $1$, most of the sites are occupied, and at large length
scales, the medium looks homogenous. One then expects that so long as $p >
p_c$, we expect the same behavior as in the system without disorder ( $p
 = 1$).

\begin{figure}
\centerline{
\epsfxsize=12.0cm
\epsfysize=9.0cm
\epsfbox{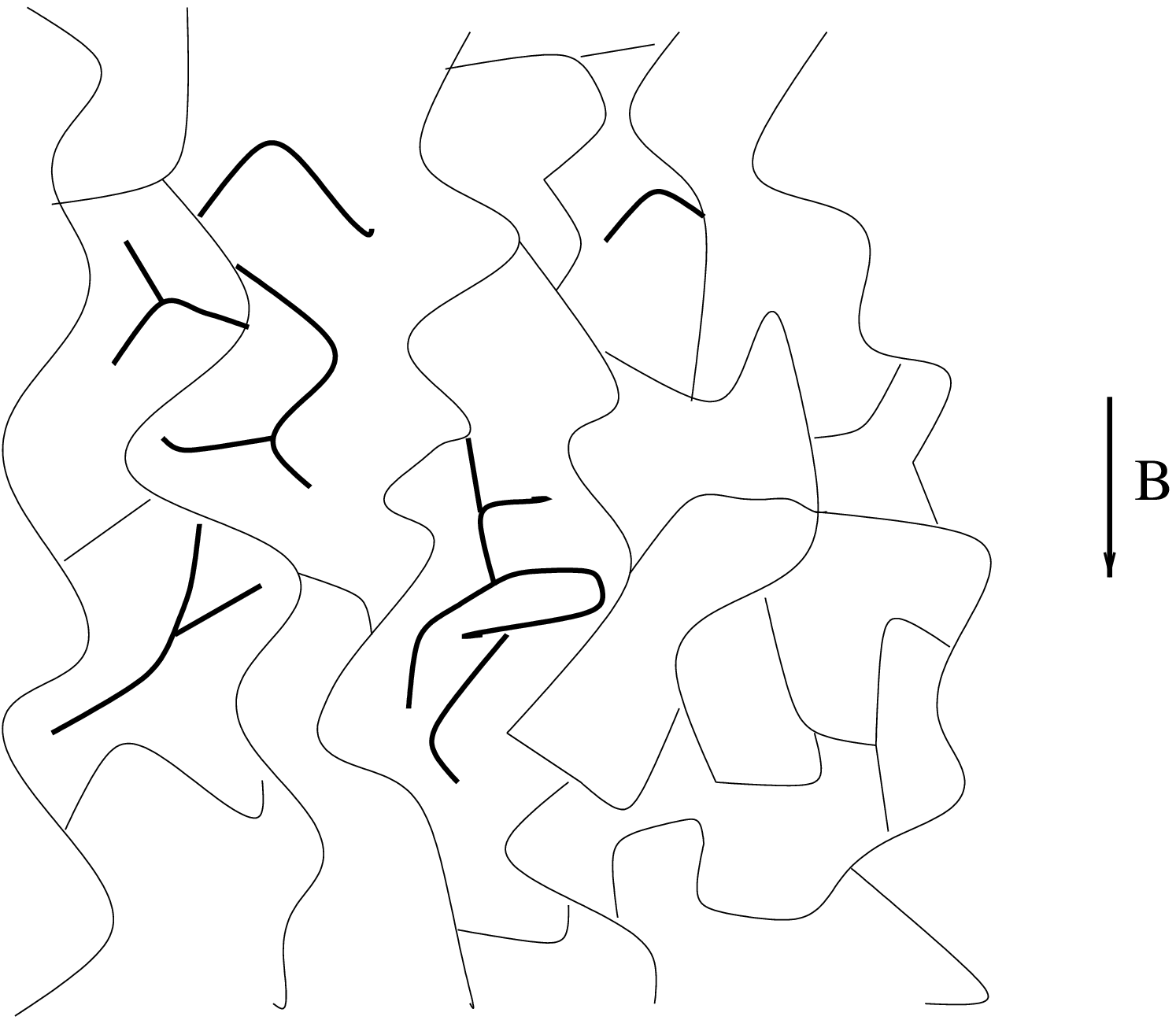}
}
\caption{}
\label{Fig.2}
\end{figure}

 Bottger and Bryksin realized that this is not so \cite{bottger}. 
They argued that the mean velocity must tend to zero as $B$ tends
to $1$, because of the possibility of trapping in  
dead-ends. We argued in \cite{barma}, that for any $p < 1$, there exists a
critical bias $B_c$ such that the asymptotic drift velocity $\vec{v}$
is exactly zero for all $B > B_c$. This is easily seen: during its motion,
the particle may get trapped in dead-end branches for long times, as it
has to move against the field to get out of the trap. For a trap of depth
$\ell$ the potential barrier to cross increases with $\ell$, and the
trapping time varies as $(\frac{1 + B}{1 - B})^{\ell}$.  The density
$\rho(\ell)$ of
traps of depth $\ell$ varies as $exp( - \ell/\xi)$. Hence the average
trapping time per step along the backbone is

\begin{equation}
\sum_{\ell=1}^{\infty} \rho(\ell)  \lgroup \frac{1 + B}{1 -
B} \rgroup^{\ell}
\label{E:traptime}
\end{equation}
where $\xi$ is the $p$-dependent percolation correlation length of the
system.
This summation converges only for $B < B_c = \tanh (\frac{1}{2 \xi})$.
For $ B = B_c - \epsilon$, with $\epsilon > 0$, this summation varies as
$1/\epsilon$, and 
the mean velocity, which varies inversely as the the mean trapping time
varies proportional to $\epsilon$.

\begin{figure}
\centerline{
\epsfxsize=12.0cm
\epsfysize=9.0cm
\epsfbox{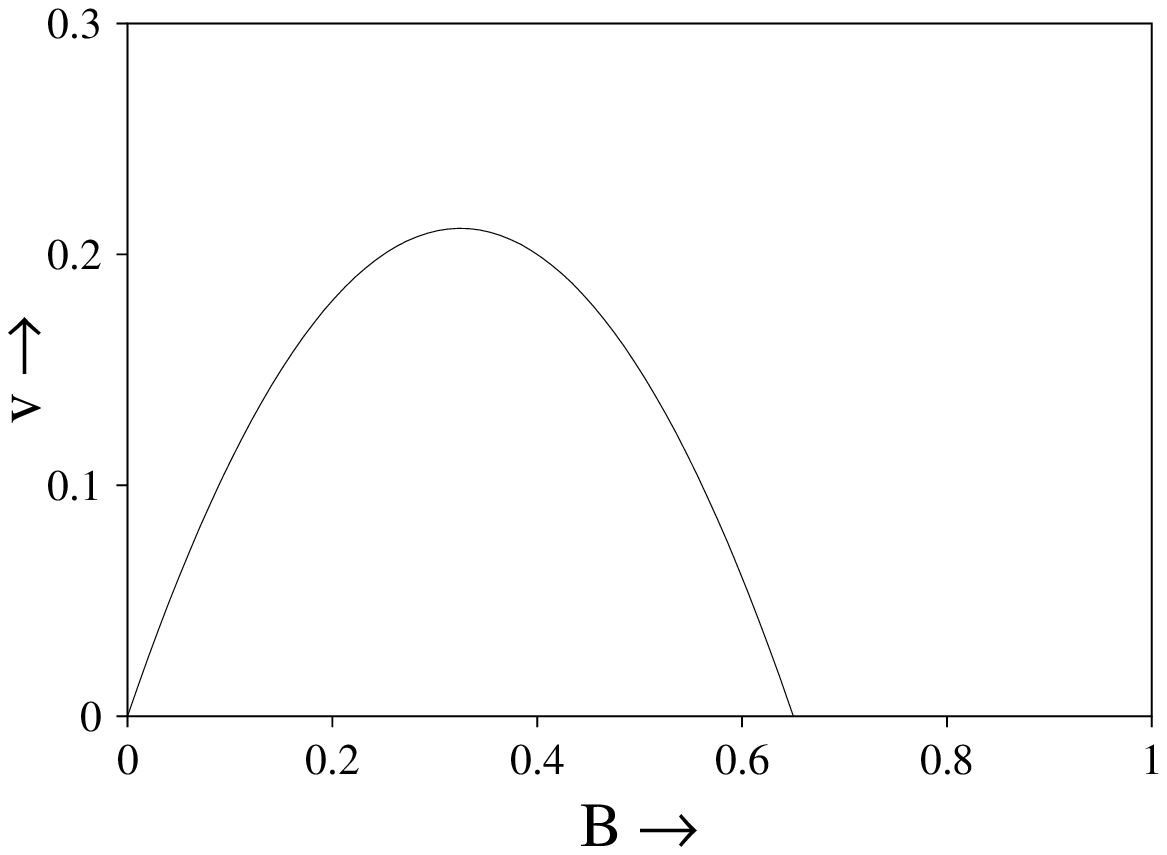}
}
\caption{}
\label{Fig.3}
\end{figure}

 For $B > B_c$, the asymptotic velocity ${\vec V}_{\infty}$ is zero, but
the mean displacement of the particle $<{\vec R}_t>$ increases as $t^a$
with $a < 1$.
The exponent $a$ depends continuously on $B$, and is easily obtained by
using the condition that the density of traps with trapping time geater
than or equal to $t$ varies as $t^{-a}$, where 
\begin{equation}
 a = (\frac{1}{\xi}) \log( \frac{ 1 + B} {1 - B}) 
\end{equation} 
In a time
$t$, the particle can, on the average, only travel a distance $t^a$ before
it encounters a trap with trapping time bigger than $t$, and gets stuck
there. Eventually, it will exit from this trap, only to get stuck in other
traps, some with an even larger trapping times. Thus, if we examine all
particles at some large time $t$, typically they would be stuck in, or
just emerging from a trap with trapping time of order $t$.

 For $B= B_c$, the average trapping time per step encountered by the
walker is given by (\ref{E:traptime}), except that summation over $\ell$
is
cutoff at a value $\ell_{max}$, where $\ell_{max}$ is the
typical value of the trapping time of the deepest trap encountered by
a walker upto time $t$. It is easy to see that $\ell_{max}$ varies as $log
t$ for large $t$, and hence the average velocity of the particle {\it up
to time} $t$ varies as $\frac {1}{log t}$.  Thus the average velocity
decreases very slowly to zero at $B = B_c$. This very slow relaxation
has
been checked in large-scale simulations  of this problem \cite{stauffer}.

 If $B = B_c - \epsilon$, then this slow decay of the velocity continues 
from its initial value of order $1$, to the final value which is of order
$\epsilon$. As the initial decay of velocity would be nearly same as that
for $B = B_c$, we see that typical relaxation time $\tau(B)$ for the
average velocity in an ensemble of non-interacting
particles to reach the steady state value varies as 

\begin{equation}
\tau(B)  \sim  exp( \frac{A}{B_c - B})
\end{equation}
where $A$ is a constant. Note that this relaxation is the overall
relaxation time for a macroscopic observable ( average current-density
for an ensemble of non-interacting particle). We are not discussing the
largest relaxation time, as that is infinite for all $B$, however small.

  Thus, biassed diffusion of noninteracting particles on a percolation
network provides a very simple model where a fast rise of relaxation time
near a dynamical phase-transition (`` the Vogel-Fulcher
 law'' of glassy dynamics) can be seen.

 We see that disorder effects both the static and dynamical properties of
the system in a very significant way. The effect on non-equilibrium
properties like response functions is much more pronounced. I hope that
further work will lead to a better understanding of these systems.

\newpage

\begin{center} {\large\underbar{\bf Captions to figures}} \end{center}
\bigskip\bigskip

\begin{enumerate} \item[{Fig. 1:}] The phase digram of a diluted magnet.
The ferromagentic, paramagnetic and Griffiths phases are denoted by PM, F,
and G respectively.

\item[{Fig. 2:}] Schematic representation of the percolation cluster with
density above the critical threshold. The heavy lines denote the dead-end
branches.

\item[{Fig. 3:}] The average velocity $v$ as a function of the biassing
field $B$.

\end{enumerate}


\newpage

\begin{thebibliography} {999}

\bibitem{percoreviews} See, for example, D. Stauffer, Phys. Rep. {\bf 54},
(1979) 1; D. Stauffer and A. Aharony, {\it Introduction to Percolation
Theory}, ( Taylor and Francis, London, 1991); G. Grimmett,{\it Percolation
Theory}, ( Springer-Verlag, NY, 1999).

\bibitem{rooted} This definition differs from the definition often used in
literature, where $\theta$ is usually defined in terms of numbers of {\it
unrooted} animals. As the number of unrooted clusters of $n$ sites is just
$A_n/n$, this definition gives a value $1$ larger than the one used by us.

\bibitem{parisi} G. Parisi and N. Sourlas, Phys. rev. Lett., {\bf 46},
(1981) 871.

\bibitem{leeyang} C. N. Yang and T. D. Lee, Phys. Rev {\bf 87}, (1952)
404; T. D. Lee and C. N. Yang, Phys. Rev. {\bf 87}, (1952) 410.

\bibitem{day} A. R. Day and T. C. Lubensky, J. Phys. {\bf A 15}, (1982)
L285; J. L. Cardy, J. Phys. {\bf A 15} (1982) 593; N. Breur and
H. K. Janssen, Z.  Phys. {\bf B 48} (1982) 347.  

\bibitem{dd83} D. Dhar, Phys. Rev. Lett.,{\bf 51} (1983) 853.

\bibitem{kunz} H. Kunz and B. Souillard, Phys. Rev. Lett. {\bf 40},(1978)
133.

\bibitem{lubensky} T. C. Lubensky and A. J. McKane, J. Phys. {\bf A 14} (
1981) L157.
 

\bibitem{griffiths} R. B. Griffiths, Phys. Rev. Lett., {\bf 23} (1969) 17.

\bibitem{ddstocha} Stochastic evolution in Ising models, in {\it
Stochastic Processes: Formalism and Applications}, Eds. G. S. Agarwal and
S. Dattagupta, Lecture Notes in Physics {\bf 184} (Springer, Berlin,
1983), p.300.

\bibitem{randeria} D. Dhar, M. Randeria and J. P. Sethna, Europhys. Lett.,
{bf 5} (1988) 485.

\bibitem{bray} A. J. Bray, Phys. Rev. Lett. {\bf 60}, (1988) 720.

\bibitem{takano} H. Takano and S. Miyashita, J. Phys. Soc. Japan, {\bf
64}, (1995) 3688.

\bibitem{unbiassed} P.G. de Gennes, La Recherche {bf 7}, 916 (1976); C. D.
Mitescu, H. Ottavi and J. Roussenq, AIP Conf. Proc. {\bf 40}, 377 (1978); 
S. Havlin and D. ben Avraham, Adv. Phys. {\bf 36}, 395 (1987).

\bibitem{bottger} H. Bottger and V. V. Bryksin, Phys. Stat. Solid. {\bf
B113}, 9 (1982).

\bibitem{barma} M. Barma and D. Dhar, J. Phys. {\bf C 16}, 1451 (1982).

\bibitem{stauffer} D. Dhar and D. Stauffer, Int. J. Mod. Phys. {\bf C 9},
349 (1998). 

\end{thebibliography}
\end{document}